# FINDING INFORMED TRADERS IN FUTURES AND THEIR UNDERLYING ASSETS IN INTRADAY TRADING[a]

*Lyudmila A. Glik* [b], *Oleg L. Kritski* [c],

*Tomsk Polytechnic University, Russia*

## Abstract

We propose a mathematical procedure for finding informed traders in ultra-high frequency trading. We wrote it as Vector ARMA and found condition of its stationarity. For the price exposure complied with ARMA(1,2) we proved that underlying asset price difference can be derived as ARMA(1,1) process. For validation of the model, we test an influence of informed traders in EUR/USD, GBP/USD, USD/RUB pairs and futures, in gold and futures prices, in Russian Trade System share index (RTS) and futures trading. We found some evidence of such influence in gold and currency pair USD/RUB pricing, in RTS index in the period from Dec 16 till Dec 20, 2013 and from Jan 28 till Jan 30.

## 1. Introduction

Identifying of insider traders, defined here as large institutional investors (such as pension or hedge funds, market makers, dealers and etc.), is an actual and complex problem. They play an important role in maintaining the operating performance of the stock market by providing it with a high liquidity, by keeping up some level of prices and trading volumes of financial instruments and by smoothing a significant price volatility. Besides the special functions, they use privileges which are not accessible to individual investors and speculators (Scholtus et al. (2014); Lepone and Yang (2013); Dey and Radhakrishna (2013)): they may obtain information about important macroeconomic indicators faster, they may also possess nonpublic information about firms before it is officially published in the quarterly or annual financial reports. For example, Scholtus et al. (2014) determine the statistical significance of return changes and show an increase in trading activity before the official announcement time of such important economic indicators as Chicago PMI and UMich Sentiment. It appears that subscribers receive information 3 and 5 minutes earlier than others. And, at the time of the announcement published data is already included in the price and significant spikes disappear.

[a] This work was financially supported by Russian Scientific Fund.
[b] Tomsk Polytechnic University, 36 Lenin Ave., Tomsk, Tomsk reg., Russia, 634050.
[c] Corresponding author. Tomsk Polytechnic University, 30 Lenin Ave., Tomsk, Tomsk reg., Russia, 634050. Tel:+73822 418913. Fax: +73822 529658.
E-mail address: olegkol@tpu.ru

We intentionally did not determine the impact of insiders on the trading activity. First, we assume that the number of insiders is low in comparison to large traders. Second, their activity is regulated, for example, by Securities Exchange Act (USA) or by Federal Law no. 224-FZ (Russia). Third, the insider may benefit from the disclosure of private information to informed traders (see, e.g., Grégoire and Huang (2012); Liu and Zhang (2011)), because it facilitates the increase of market activity and stock prices, therefore profit of the insider increases too.

The literature on the detection of insider and informed traders is large. Even so, special attention is devoted to the classic paper of Kyle (1985). It employs basic ideas and an original two-period Kyle model for insider trading using a linear price disclosure function and existence of the equilibrium strategy with no public disclosure, considering the market efficiency and profit maximization by the monopolistic insider. Later, the approach was commonly accepted and developed in many works (Grégoire and Huang (2012); Liu and Zhang (2011); Wang et al. (2009); Daher et. al. (2012)). So, Grégoire and Huang (2012); Liu and Zhang (2011) developed mathematical models of interaction between one insider, $N$ institutional investors, a number of individual ("noise") traders and one market maker. In Daher et. al. (2012) authors implied the Cournot competition (duopoly) in the Kyle model and, by analogy with Wang et al. (2009), examined the Stackelberg competition where insiders are the owner of the company and top manager subordinated to him.

Ideas of Kyle influenced the papers of Deb et al. (2014); Zhou (2012); Vitale (2012); Piccione and Spiegler (2014); Yang and Li (2014)). So, Deb et al. (2014) examines the competition among informed traders and a market maker in the presence of a noise trader. The authors establish a three-period rational expectations equilibrium (REE) model where informed traders receive and verify a noisy signal, and trade by their own.

In Zhou (2012) the original two-period Kyle model is generalized for the case of a single insider, who has some confidence degree, multiple market makers and multiple noise traders. Authors find the optimal trading strategies for such insider, based on the information available to him and his belief level (confident or pessimistic). Using the same initial assumptions Vitale (2012) presumes that such insider is risk-averse, implying a multi-security version of Kyle's model with Markovian linear sequential equilibrium.

Lastly, papers of Piccione and Spiegler (2014); Yang and Li (2014)) suggest a trading model with investor sentiment that extends the original Kyle model. In this model, there are an informed rational investor, who maximizes his expected profits, and uninformed 'noise' sentiment speculator, who trades on sentiment shocks. Authors derive an analytical solution to the sentiment equilibrium prices and describe a dynamic price path.

There exists a significant disadvantage in all the foregoing papers – they employ the linearity of the risky asset price exposure, which in our opinion, limits the scope of their applicability.

The paper of Park and Lee (2010) presents a nonlinear model of the dynamics of risky asset returns in the presence of insider trading. The authors disclose the dependence of risky asset returns on the amount of insider transactions and denote it in the form of an ARMA(1,1) model. Furthermore, they obtain the detection criteria of insider trading which proved to be closely related with stability conditions of the ARMA(1,1) process.

Along with the identification of insiders and informed traders only by changes in underlying asset prices there exist methods that involve its derivatives, notably options (Popescu and Kumar (2013); Hu (2013); Muravyev et al. (2013)). Thus, Popescu and Kumar (2013) verify the put-call parity for American options with two different strikes for different market agents and compute the probability of informed trading (PIN). And, Hu (2013) found an order imbalance in trades of the underlying asset and its three options, i.e. in-the-money (ITM), at-the-money (ATM) и out-of-the-money (OTM) options, that enables to compute the probability of informed traders. Muravyev et al. (2013) combine those two approaches.

The possibility of addressing futures prices for detection of informed traders was first noted in Yi-Tsung (2013), where it is shown that the enhancement of the trading activity in the futures market anticipates an increased activity in the spot-market, but in most cases the direction of trade (to sell or to buy) remains unclear, even if using the non-public information.

The purpose of this study is to develop the methodology for identification of the impact of informed traders on the trading activity considered in Park and Lee (2010). Hence, we present a Vector ARMA model for tick prices of futures and its underlying asset and estimate limits of its instability. Furthermore, we formulate relevant decisive criteria. The computational algorithm is applied for the analysis of high frequency data for the period that precedes announcements of important macroeconomic information about phasing out the program of "quantitative easing" (QE3) and U.S. GDP growth for the third quarter of 2013, as well as in the period following this news.

This paper is organized as follows. In Section 2 we construct a mixed multi-period model of informed trading in terms of stock and futures return series and investigate its characteristics. In Section 3 we analyze the effectiveness of detection criteria through a simulation and apply it to FX pairs, spot gold prices, developing country share index and their futures in time of QE3 ending announcements. Finally, Section 4 presents our conclusions.

## 2. Model

Let's assume that number of all players on the stock market, trading an underlying asset and its futures contract, are divided into informed traders and ordinary "noise" traders. Let a macroeconomic announcement affecting the price, be publicly known at a future time T, while the informed trader has obtained the data at time $t<T$. We assume that he decides to buy (sell) the underlying asset or futures in equal installments at regular intervals, i.e. at time $t$, $(t+1)$, …, $T$. Then the effect of changes in the underlying asset price could be defined as

$$X_t = v_t + u_t,$$

where $u_t \sim N(0, \sigma_u^2)$ – addition to the price, offered by uninformed traders, $v_t$ – surcharge to the price that an informed trader is willing to pay.

Let $v_t$ obey the relation

$$v_t = \beta \theta_t,$$

where $\beta$ – coefficient of proportionality, $\theta_t$ – value of trading transaction of informed trader.

We suppose that $\theta_t$ satisfies AR(1) model, which is explained by the desire of the informed trader to hide his activity and, for instance, to reduce his contribution $v_t$ at low levels of market activity:

$$\theta_t = \overline{\theta} + \rho \theta_{t-1} + z_t, \tag{1}$$

where $\overline{\theta}$ – average value of trading transaction price per time unit, $z_t \sim N(0, \sigma_z^2)$ – price noise.

Let $S_t$ be a quote of the underlying asset at time $t$. Since a trader buys a large quantity of the underlying asset, we assume that $S_t$ will vary proportionally to the change of prices:

$$S_t = S_{t-1} + \lambda X_t, \tag{2}$$

where $\lambda = \dfrac{\mathrm{cov}((\theta_t, X_t)|v_{t-1})}{D(X_t|v_{t-1})}$ – market depth variation coefficient.

It follows from equation (2) that the coefficient $\theta_t$ defines a price exposure with respect to all market participants. Moreover in our case it is easy to verify that

$$\lambda = \frac{\beta \sigma_z^2}{\beta \sigma_z^2 + \sigma_u^2},$$

If noise processes $z_t, u_t$ are independent of one another.

Using an analytical representation of $X_t$ and substituting $\theta_t$ defined by (1), equation (2) can be written as an ARMA (1,1) model from Park and Lee (2010):

$$\Delta S_{t+1} = \gamma + \rho \Delta S_t + \delta \varepsilon_t + \varepsilon_{t+1}, \qquad (3)$$

where $\varepsilon_t \sim N(0, \sigma_\varepsilon^2)$ is a noise process, $\sigma_\varepsilon^2 = \left(\lambda^2 \beta^2 \sigma_z^2 + (1-\rho^2)\lambda^2 \sigma_u^2\right)\left(1+\delta^2+2\rho\delta\right)^{-1}$,

$\delta = (1+\rho^2)(2\rho)^{-1} + \left[\lambda^2\beta^2\sigma_z^2 - \sqrt{\left(\lambda^2\sigma_u^2(1-\rho)^2 + \lambda^2\beta^2\sigma_z^2\right)\left(\lambda^2\sigma_u^2(1+\rho)^2 + \lambda^2\beta^2\sigma_z^2\right)}\right]\left(2\rho\lambda^2\sigma_u^2\right)^{-1}$,

$\gamma = \lambda\beta(1-\rho)\dfrac{S_T - S_0}{T}$.

Representation of equation (2) in the form of (3) allows us to entail the following conditions from Park and Lee (2010): 1) if ρ<0, then 0<δ<-ρ; 2) if ρ>0, then -1<δ<-ρ. We could, therefore, formulate the following criterion:

**Criterion**: coefficients of the AR(1) and MA(1) parts of model (3) (ρ and δ, respectively) must have opposite signs. If $\rho > 0$, then based on the stability conditions, $\rho < |\delta|$. If $\rho < 0$, then $|\rho| > \delta$.

Let $S_t$, $t = 0,1,..,T$, be a data set that is available for analysis. Let $m<T$ be a time window length, which allows us to calculate the initial estimates of coefficients $\hat{\gamma}_1 = \left(\hat{\lambda}_1 \beta(1-\hat{\rho}_1)\dfrac{S_m - S_0}{m}\right)$, $\hat{\rho}_1$ and $\hat{\delta}_1$ with quotes $S_0, S_1, S_2, ..., S_m$ in model (3). Moving the time window to the right per unit until we reach time T, by known $S_s, S_{1+s}, ..., S_{m+s}$ we estimate $\hat{\gamma}_s$, $\hat{\rho}_s$ and $\hat{\delta}_s$, $s = 0,1,..,(T-m)$. Furthermore, we use empirical values of the coefficients found to formulate a decision rule. Accordingly, we write the criterion:

**Decisive criterion**: subject to non-zero price movements $\hat{\gamma}_k \neq 0$, $k = 0,1,..,(T-m)$, we assume that the informed transaction is detected if one of the following inequalities holds:

a) $\sum_k \hat{\rho}_k \sum_k \hat{\delta}_k < 0$, $\sum_k \hat{\rho}_k < 0$, $\left|\sum_k \hat{\rho}_k\right| > \left|\sum_k \hat{\delta}_k\right|$; б) $\sum_k \hat{\rho}_k \sum_k \hat{\delta}_k < 0$, $\sum_k \hat{\rho}_k > 0$, $\left|\sum_k \hat{\rho}_k\right| < \left|\sum_k \hat{\delta}_k\right|$.

Further, let $F_t$ be a futures price of the underlying asset priced at $S_t$. It is well known that there exists a relation (Alexander and Barbosa (2008)):

$$F_t = S_t \exp(r(T-t)), \qquad (4)$$

where $r$ – risk-free interest rate, $T$ – expiry time for a futures contract, $t$ – current time, one can write the analogue of equation (3) for returns $\Delta F_{t+1}$.

Now we state our main results of this paper.

**Theorem 1.** *Let's suppose* $A_{t+1} = \left[\gamma + (1+\rho - e^r - \rho e^{-r})S_0\right]\exp(r(T-t-1))$, $\tilde{\varepsilon}_t = \varepsilon_t \exp(r(T-t-1))$ *is a Gaussian noise process*, $C_{t+1} = (1+\rho - e^r - \rho e^{-r})\exp(r(T-t-1))$. *Then*,

$$\Delta F_{t+1} = A_{t+1} + \rho e^{-2r}\Delta F_t + C_{t+1}\sum_{j=1}^{t}\Delta S_j + \delta\tilde{\varepsilon}_t + \tilde{\varepsilon}_{t+1}. \qquad (5)$$

The proof of theorem 1 is listed in *Appendix A*.

**Theorem 2.** *System of equalities (3) and (5) can be written in the form of the Vector ARMA model*:

$$X_{t+1} = \tilde{A} + X_t\tilde{B} + \sum_{j=1}^{t-1}X_{t-j}\tilde{C} + \delta(\tilde{\varepsilon}_t;\varepsilon_t) + (\tilde{\varepsilon}_{t+1};\varepsilon_{t+1}), \qquad (6)$$

where $X_{t+1} = (\Delta F_{t+1}; \Delta S_{t+1})$, $\tilde{A} = (A_{t+1};\gamma)$ – vectors, $\tilde{B} = \begin{pmatrix} \rho e^{-2r} & 0 \\ C_{t+1} & \rho \end{pmatrix}$, $\tilde{C} = \begin{pmatrix} 0 & 0 \\ C_{t+1} & 0 \end{pmatrix}$ – square matrices 2×2.

The proof of Theorem 2 is obvious.

We note that representing the process in the form of (6) allow us to determine the conditions of its stationarity (see, e.g., Wei (2006) for getting conditions and estimation of the stationarity of Vector ARMA processes). From the results we can formulate Theorem 3:

**Theorem 3.** *Process $X_{t+1}$ in equation (6) is stationary, if $|\rho|<1$.*

The proof of theorem 3 is listed in *Appendix A*.

Theorem 3 enables us to write the necessary generalized criterion of the presence of informed traders when trading futures contracts and the underlying asset on them:

**Generalized criterion**: 1) if -1<ρ<0, then 0<δ<-ρ; 2) if 1>ρ>0, then -1<δ<-ρ.

The decisive criterion will barely change and can be rewritten easily.

At the end of this section we notice that if unlike equation (1) we choose the price exposure process $\theta_t$ as ARMA(1,2), i.e.

$$\theta_t = \bar{\theta} + \rho\theta_{t-1} + z_t + z_{t-1}, \qquad (7)$$

then equation (3) for underlying asset price should be rewritten under the following theorem.

**Theorem 4.** *In case an informed trader implements a mixed strategy (2) with $\lambda = \dfrac{4\beta\sigma_z^2}{4\beta\sigma_z^2 + \sigma_u^2}$ as the information exposure strategy in a multi-period model, underlying asset price difference follows the ARMA(1,1) process below.*

$$\Delta S_t = \gamma + \rho\Delta S_{t-1} + \varepsilon_t + \delta\varepsilon_{t-1}, \qquad (8)$$

where $\varepsilon_t \sim N(0, \sigma_\varepsilon^2)$, $\sigma_\varepsilon^2 = \lambda^2 \beta^2 \sigma_z^2 (1+\rho^2)(\rho\delta^2 + \rho^2\delta + \rho + \delta)^{-1}$, $\gamma = \lambda\beta(1-\rho)\frac{S_T - S_0}{T}$,

$\delta = \left[\sigma_u^2(1+\rho^2) + 2\beta^2\sigma_z^2 - (1+\rho)\sigma_u\sqrt{4\beta^2\sigma_z^2 + \sigma_u^2(1-\rho)^2}\right]\left(2\rho\sigma_u^2 - 2\beta^2\sigma_z^2\right)^{-1}$.

The proof is listed in the *Appendix A*.

## 3. Numerical simulation results

### 3.1. Data

We use ultra-high frequency tick data for EUR/USD, GBR/USD and USD/RUB currency pairs and spot gold prices, their March, 2014 futures, 1 min data for Russian Trade System share index (RTS) and its futures (ticker RIH4) in the period Dec 17 – Dec 20, 2013 and Jan 28 – Jan 30, 2014. The data were obtained from FINAM holding (www.finam.ru) and CME Group.

The choice of financial instruments for analysis is due to macroeconomic statistics, its information is disclosed in certain periods of time: it is news about phasing out the U.S. monetary stimulus, i.e. QE3, and final data on U.S. GDP for the III quarter of 2013, and preliminary - for the IV quarter of 2013, from which the termination time of QE3 was depended on.

### 3.2. Results

While three incentive programs were functioning, the first one was launched in March, 2009, the role of regional currencies (Australian and New Zealand dollar, Brazilian and Mexican peso, Chinese yuan, South African rand, Russian ruble, Turkish lira and Indonesian rupiah) significantly increased (Triennial Central Bank Survey Report (2013)): for example, the percentage share of rouble in "net-net" turnover in the FX market increased from 0,9% to 1,6% for last three years. And this situation is not unique: the share of RMB increased over the same period from 0.9% to 2.2%, the Mexican peso - from 1.3% to 2.5%, the Turkish lira - from 0.7% to 1.3 %. These facts indicate that there was a free cash flow from developed countries to developing ones over the past three years. On the one hand, it stimulated their economies and strengthened regional currencies up to a certain point of time. On the other hand, imbalances increased and overheating of economies occurred at the first threat of collapse of the incentive programs, which led to a sharp devaluation of the local currency and the stock market crashes. All of these help to identify large informed traders and fix their activity, for example, in the Russian or any other emerging markets.

The possible existence of relationship between FX market and gold prices during the crisis is discussed in the paper by Pukthuanthong and Roll (2011), where strong positive correlations between gold prices and Dollar, Euro, Yen and Pound were found and investigated. Therefore, we investigate the activity of traders in trading of gold futures and examine the gold spot prices.

Let's define the impact of informed traders in intraday trading of currency pairs EUR/USD, GBR/USD, USD/RUB and March, 2014 futures.

The results of calculations with the time window length $m=1$ hour, are presented in Table 1, *Panel A – C*.

[Table 1 is here]

According to the results of calculations, given in Table 1, Panel A-C we conclude that there is no significant impact of informed traders in trading of highly liquid currency pairs EUR/USD and GBR/USD, while for emerging market currencies USD/RUB we traced it on January 29, 2014, due to the observed capital flight from emerging to developed markets while phasing out QE3.

Because of the relationship of FX market and gold prices discussed in Pukthuanthong and Roll (2011), it is interesting to investigate the activity of traders in the gold trade at the beginning of the global structural changes.

We analyze the tick retail price of gold and futures on it. It is known that these prices are closely associated with capital inflows to developing countries and affect their foreign exchange rates (Apergis (2014)), inflation and FX reserves. The results of calculations with the width of the time window m = 1h present in Tables. 2. The results of calculations with the time window length $m=1$ hour, are presented in Table 2.

[Table 2 is here]

According to the results of calculations, given in Table 2 we found the impact of informed traders in gold trading in days following announcements of phasing out the program QE3. It is predictable since the termination of QE3 has a direct impact on expectations of large players about the reduction of supply of dollar in the world and, as a consequence, the decrease of precious metal prices.

In our opinion, such reason is based on the drop of indices in developing countries. We show it on the example of an emerging stock market, such as Russia.

We analyze 1-minute data of March, 2014 futures on RTS index (ticker RIH4) priced in rouble and 1-minute data of RTS index priced in dollar. These index assets have a rouble-dollar nature and the quotes depend directly on EUR/USD rate associated with processes of capital flight.

The results of calculations with the time window length $m=1$ hour, are presented in Table 3.

[Table 3 is here]

Thus, our calculations identify two different stages after the announcement of the termination of QE3. At the first stage (Dec 18 - Dec 20, 2013) informed traders showed high activity in gold trading and futures on it, but closer to the end of the stage they switched to index assets of emerging markets that our calculations revealed. At the second stage (Jan 28 - Jan 30, 2014) operations with currencies of developing countries (e.g., RUB) were added to the trading activity that is correlated to the capital flight to developed markets. At the same time suspicious transactions with highly liquid pairs EUR/USD, GBR/USD were not detected in the FX market, despite the high volatility of their prices throughout the first stage.

**4. Conclusion**

Our analysis shows that the proposed generalized and decisive criteria allow to identify the activity of informed traders in intraday trading in global markets. Further, it is shown that such traders while are unable to even in the short term to set their movement trajectory prices on all stock markets, i.e. principles of fair and honest trading are fulfilled. Further, we reveal the current inability of such traders to determine price trajectories in all stock markets, i.e. principles of fair and honest trading are fulfilled. At the same time, informed trading in the emerging market of Russia is detected easier while using a wide range of financial instruments.

*APPENDIX A*. **Proofs.**

To prove Theorem 1, we first state and prove an auxiliary Lemma 1.

**Lemma 1**: If $F_t$ holds equality (4), then

$$F_{t+1} = F_t e^{-r} \frac{S_{t+1}}{S_t}.$$

*Proof.* By (4), $F_t = S_t \exp(r(T-t))$, $F_{t+1} = S_{t+1} \exp(r(T-t-1))$. Hence, $\ln \frac{F_{t+1}}{F_t} = \ln \frac{S_{t+1}}{S_t} - \ln e^r$.

*Proof of Theorem 1.* According to (3), $\Delta S_{t+1} = \gamma + \rho \Delta S_t + \delta \varepsilon_t + \varepsilon_{t+1}$, or, in terms of prices, $S_{t+1} = \gamma + (1+\rho)S_t - \rho S_{t-1} + \delta \varepsilon_t + \varepsilon_{t+1}$. We apply this expression to the results in the proof of Lemma 1. Then,

$$F_{t+1} = F_t e^{-r} \frac{\gamma + (1+\rho)S_t - \rho S_{t-1} + \delta \varepsilon_t + \varepsilon_{t+1}}{S_t}, \text{ or}$$

$$F_{t+1} = F_t e^{-r} \left[ \frac{\gamma + \delta \varepsilon_t + \varepsilon_{t+1}}{S_t} + (1+\rho) - \rho \frac{S_{t-1}}{S_t} \right].$$

Let's use Lemma 1 once again: $\frac{S_{t-1}}{S_t} = \left(\frac{S_t}{S_{t-1}}\right)^{-1} = \left(\frac{F_t e^r}{F_{t-1}}\right)^{-1} = \frac{F_{t-1}}{e^r F_t}$. So

$$F_{t+1} = F_t e^{-r} \left[ \frac{\gamma + \delta\varepsilon_t + \varepsilon_{t+1}}{S_t} + (1+\rho) - \rho \frac{F_{t-1}}{e^r F_t} \right],$$

or, by (4), $e^r F_{t+1} = F_t \left[ \frac{\gamma + \delta\varepsilon_t + \varepsilon_{t+1}}{F_t} \exp(r(T-t)) + (1+\rho) - \rho \frac{F_{t-1}}{e^r F_t} \right]$, i.e.

$$e^r F_{t+1} = (\gamma + \delta\varepsilon_t + \varepsilon_{t+1}) \exp(r(T-t)) + (1+\rho) F_t - \rho \frac{F_{t-1}}{e^r}.$$

Let's form the difference $\Delta F_{t+1}$ in the left side of the equation, then we subtract and add $e^r F_t$:

$$e^r \Delta F_{t+1} = (\gamma + \delta\varepsilon_t + \varepsilon_{t+1}) \exp(r(T-t)) + (1+\rho - e^r) F_t - \rho \frac{F_{t-1}}{e^r}.$$

We proceed similarly with the last term on the right side of the equation:
now,

$$-\rho F_{t-1} e^{-r} = \rho \Delta F_t e^{-r} - \rho e^{-r} F_t,$$

then

$$e^r \Delta F_{t+1} = (\gamma + \delta\varepsilon_t + \varepsilon_{t+1}) \exp(r(T-t)) + (1+\rho - e^r - \rho e^{-r}) F_t + \rho e^{-r} \Delta F_t.$$

Let's transform $F_t$. substituting the price equation in (4) with $S_t = \Delta S_t + S_{t-1} = \Delta S_t + \Delta S_{t-1} + S_{t-2} = \ldots = \sum_{j=1}^{t} \Delta S_j + S_0$, where $S_0$ – known initial price of the underlying asset, and dividing both sides of the equation by $e^r$, considering notions of the theorem we obtain equation (5).

*Proof of Theorem 3.* According to the general theory of stationary of Vector ARMA processes (Wei (2006)), $X_{t+1}$ is stationary if and only if all eigenvalues of $\tilde{B}$ and $\tilde{C}$ in (6) lie inside the unit circle in the complex plane. Clearly, matrix $\tilde{B}$ has two real eigenvalues: $\rho e^{-2r}$ and $\rho$, and matrix $\tilde{C}$ – zero eigenvalue of multiplicity 2. Since $|\rho| e^{-2r} < |\rho| < 1$ by condition of the theorem, then $X_{t+1}$ is stationary.

*Proof of Theorem 4.* After double substitution of equation 1 into equation 2 with different $t$ we obtain

$$\Delta S_t = \gamma + \rho\lambda\beta\theta_{t-1} + \lambda\beta z_t + \lambda\beta z_{t-1} + \lambda u_t, \tag{A.1}$$

$$\Delta S_{t+1} = \gamma(1+\rho) + \rho^2\lambda\beta\theta_{t-1} + \lambda\beta(1+\rho) z_t + \lambda\beta\rho z_{t-1} + \lambda\beta z_{t+1} + \lambda u_{t+1}. \tag{A.2}$$

We modify (A.2) as

$$\Delta S_{t+1} = \gamma + \rho\Delta S_t + \lambda\beta z_{t+1} + \lambda\beta z_t + \lambda u_{t+1} - \lambda\rho u_t. \tag{A.3}$$

We denote the autocovariance of some auxiliar expression

$$\Delta S_t = \gamma + \rho\Delta S_{t-1} + \varepsilon_t + \delta\varepsilon_{t-1}$$

as $V_j$, so $V_0$ and $V_1$ can be easily computed as follows:

$$V_0 = \sigma_\varepsilon^2 (1+\delta^2 + 2\rho\delta)(1-\rho^2)^{-1}, \tag{A.4}$$

$$V_0 = \sigma_\varepsilon^2 (\rho + \rho\delta^2 + \rho^2\delta + \delta)(1-\rho^2)^{-1}. \tag{A.5}$$

If we compute similar autocovariances of (A.3) directly and equate them to (A.4) and (A.5), we can find unknown $\delta$ and $\sigma_\varepsilon^2$. The system of such equations is as written below:

$$\sigma_\varepsilon^2 (1+\delta^2 + 2\rho\delta)(1-\rho^2)^{-1} = \lambda^2 \beta^2 \sigma_z^2 (1-\rho)^{-1} + \lambda^2 \sigma_u^2,$$

$$\sigma_\varepsilon^2 (\rho + \rho\delta^2 + \rho^2\delta + \delta)(1-\rho^2)^{-1} = (1+\rho)\lambda^2 \beta^2 \sigma_z^2 (1-\rho)^{-1}.$$

So, it is proved that

$$\delta = \left[\sigma_u^2 (1+\rho^2) + 2\beta^2 \sigma_z^2 - (1+\rho)\sigma_u \sqrt{4\beta^2 \sigma_z^2 + \sigma_u^2 (1-\rho)^2}\right] \left(2\rho\sigma_u^2 - 2\beta^2 \sigma_z^2\right)^{-1},$$

$$\sigma_\varepsilon^2 = \lambda^2 \beta^2 \sigma_z^2 (1+\rho^2)(\rho\delta^2 + \rho^2\delta + \rho + \delta)^{-1}.$$

These coefficients, obviously, define our auxiliar expression in the form (8) needed.

Table 1

Impact of informed traders on trading process in the analysis of EUR/USD, GBR/USD, USD/RUB and their March, 2014 futures tick prices at different dates

*Panel A. EUR/USD*

| Date | Underlying asset | | Futures | | Presence of impact |
|---|---|---|---|---|---|
| | $\bar{\rho}$ | $\bar{\delta}$ | $\bar{\rho}$ | $\bar{\delta}$ | |
| 17.12.2013 | 0,176 | 0,327 | 0,168 | 0,277 | declined |
| 18.12.2013 | 0,769 | 0,836 | 0,309 | 0,679 | declined |
| 19.12.2013 | 0,344 | 0,443 | 0,869 | 0,908 | declined |
| 20.12.2013 | 0,244 | 0,350 | -0,639 | -0,667 | declined |
| 28.01.2014 | 0,008 | 0,219 | 0,507 | 0,559 | declined |
| 29.01.2014 | -0,079 | 0,082 | 0,058 | 0,152 | declined |
| 30.01.2014 | 0,788 | 0,848 | -0,894 | -0,866 | declined |

*Panel B. GBR/USD*

| Date | $\bar{\rho}$ | $\bar{\delta}$ | $\bar{\rho}$ | $\bar{\delta}$ | Presence of impact |
|---|---|---|---|---|---|
| 17.12.2013 | 0,740 | 0,769 | -0,091 | -0,196 | declined |
| 18.12.2013 | 0,034 | 0,099 | 0,278 | 0,374 | declined |
| 19.12.2013 | -0,007 | 0,089 | -0,643 | -0,689 | declined |
| 20.12.2013 | -0,457 | -0,403 | 0,347 | 0,286 | declined |
| 28.01.2014 | 0,782 | 0,808 | 0,361 | 0,713 | declined |
| 29.01.2014 | 0,057 | 0,206 | 0,361 | 0,662 | declined |
| 30.01.2014 | 0,598 | 0,656 | -0,090 | -0,208 | declined |

*Panel C. USD/RUB*

| Date | $\bar{\rho}$ | $\bar{\delta}$ | $\bar{\rho}$ | $\bar{\delta}$ | Presence of impact |
|---|---|---|---|---|---|
| 17.12.2013 | -0,120 | 0,746 | -0,112 | -0,193 | declined |
| 18.12.2013 | 0,090 | 0,700 | 0,380 | 0,394 | declined |
| 19.12.2013 | -0,020 | 0,781 | -0,503 | -0,619 | declined |
| 20.12.2013 | -0,065 | 0.789 | 0,247 | 0,295 | declined |
| 28.01.2014 | -0,104 | 0,581 | -0,211 | -0,312 | declined |
| 29.01.2014 | -0,696 | 0,196 | -0,582 | 0,275 | accepted |
| 30.01.2014 | -0,232 | 0,431 | -0,210 | 0,331 | declined |

Table 2

Impact of informed traders on trading process in the analysis of tick spot gold prices and March, 2014 futures tick prices at different dates

| Date | Gold | Futures | Presence of impact |
|---|---|---|---|

|            | $\hat{\rho}$ | $\hat{\delta}$ | $\hat{\rho}$ | $\hat{\delta}$ |          |
|------------|--------|--------|--------|--------|----------|
| 17.12.2013 | -0,115 | 0,231  | -0,055 | 0,138  | declined |
| 18.12.2013 | -0,315 | 0,166  | -0,298 | 0,061  | accepted |
| 19.12.2013 | -0,077 | 0,289  | -0,086 | 0,233  | declined |
| 20.12.2013 | -0,087 | 0,244  | -0,013 | 0,202  | declined |
| 28.01.2014 | -0,034 | 0,279  | 0,462  | 0,782  | declined |
| 29.01.2014 | -0,251 | 0,050  | 0,334  | 0,709  | accepted |
| 30.01.2014 | 0,144  | 0,299  | 0,454  | 0,763  | declined |

Table 3

Impact of informed traders on trading process in the analysis of 1-minute data of RTS index and prices of RIH4 futures at different dates

| Date | RTS Index | | Futures on RTS Index | | Presence of impact |
|------|-----------|---|----------------------|---|--------------------|
|      | $\hat{\rho}$ | $\hat{\delta}$ | $\hat{\rho}$ | $\hat{\delta}$ | |
| 17.12.2013 | -0,655 | -0,695 | 0,251  | 0,352  | declined |
| 18.12.2013 | -0,853 | -0,756 | 0,178  | 0,299  | declined |
| 19.12.2013 | -0,633 | -0,688 | -0,114 | -0,039 | declined |
| 20.12.2013 | 0,808  | 0,773  | -0,146 | 0,017  | accepted |
| 28.01.2014 | 0,874  | 0,845  | -0,306 | -0,295 | declined |
| 29.01.2014 | 0,205  | 0,043  | 0,769  | 0,724  | declined |
| 30.01.2014 | 0,258  | 0,046  | -0,431 | -0,439 | declined |